\documentclass[12pt]{article}

\usepackage{amssymb}
\usepackage{amsfonts}

\usepackage{pstricks}

\begin{document}

\title{Landau--Zener formula and non-secular transitions}

\author{S.V.Kozyrev\footnote{Steklov Mathematical Institute, Moscow, Russia}}

\maketitle

\begin{abstract}
The relationship between the Landau--Zener model (which describes transitions between energy levels for a time-dependent Hamiltonian) and non-secular master equations is discussed. This approach allows to describe the widely discussed in the literature on quantum photosynthesis resonance of the energy of the vibron and the difference of exciton energies for transitions between exciton states coupled to the vibron.
\end{abstract}

\section{Introduction}

In this paper, we discuss the connection between Landau--Zener transitions and master equations with non-secular terms. Landau--Zener formula \cite{Landau}, \cite{Zener}, \cite{Stuckelberg}, \cite{Majorana} describes the probabilities of transitions between energy levels for a time-dependent Hamiltonian. Physically, this can be a model of transitions between electronic states in a molecule generated by vibrations of nuclei. Transitions happen when the oscillation passes the avoided crossing point for energy levels. If we consider a slow time scale, large compared to the period of vibronic oscillations, transitions between energy levels will be described by the von Neumann equation for density matrix, and the Hamiltonian in this equation will contain a non-secular term and will be different from the original Hamiltonian of the system. Taking into account decoherence and dissipation will add a dissipative GKSL generator to the equation, as a result, the dynamics of the system will be described by a master equation with non-secular terms.

This way of appearance of non-secular terms in master equations can be considered as a mechanism for amplification of transport of electronic excitations in quantum photosynthesis. In models of quantum photosynthesis, it was proposed to consider non-secular terms of ''laser--like type'' \cite{Chin2013}. In \cite{laser} a similar model has been discussed as a quantum transfer amplifier. In quantum photosynthesis, the appearance of such terms is explained not by pumping as in laser models, see \cite{Haken} for exposition of theory of lasers, but by presence of vibrons  --- special modes of vibration of nuclei in molecules. Vibrons are generated with excitation of electronic states according to the Franck--Condon principle --- upon exciton excitation, the nuclei are in a nonequilibrium state and begin to vibrate. This causes transitions between electronic states according to the Landau--Zener mechanism. These transitions can be seen as generated by non-secular terms in the master equation.

The presentation of the paper is organized as follows. In Section 2, we discuss the Landau--Zener formula and present an illustrative non-rigorous derivation of this formula. Section 3 considers non-secular master equations related to vibrons and Landau--Zener transitions.

\section{Landau--Zener formula}

\noindent{\bf Transition between potential wells}.
Let us discuss the transition between electronic states in a molecule of the kind of tunneling between potential wells. The corresponding Hamiltonian has the form
\begin{equation}\label{Hamiltonian}
H=\pmatrix{E_1 & J \cr J & E_2 \cr}.
\end{equation}
If the energies of the wells are equal, i.e. $E_1=E_2=E$, then eigenstates of the Hamiltonian are
$$
\psi_{+}=\frac{|1\rangle+|2\rangle}{\sqrt{2}},\quad \psi_{-}=\frac{|1\rangle-|2\rangle}{\sqrt{2}},\quad |1\rangle=\pmatrix{ 1 \cr 0 },\quad |2\rangle=\pmatrix{ 0 \cr 1 }
$$
and eigenvalues are $E_{\pm}=E\pm J$.

Let us consider the Schroedinger evolution
$$
\frac{\partial}{\partial t}\psi(t)=-iH\psi(t),
$$
$$
e^{-itH}|1\rangle=\frac{1}{\sqrt{2}}e^{-itE_{+}}\left(\psi_{+}+e^{2itJ}\psi_{-}\right),
$$
we get for the transition amplitude
\begin{equation}\label{Amplitude0}
\langle 2|e^{-itH}|1\rangle=\langle\frac{\psi_{+}-\psi_{-}}{\sqrt{2}},e^{-itH}\frac{\psi_{+}+\psi_{-}}{\sqrt{2}}\rangle=\frac{1}{2}e^{-itE_{+}}\left(1-e^{2itJ}\right).
\end{equation}

In this system quantum beats will be observed --- if we start from $|1\rangle$ then after the period $T=\frac{\pi}{2J}$ we will get $|2\rangle$. Velocity of the transition is determined by the level splitting $2J$.

Presence of energy difference for potential wells interferes with tunneling between these wells. Let energy levels be non-equal $E_1\ne E_2$, we denote $E_1+E_{2}=2E$, $E_1-E_{2}=2\Delta$ (let $\Delta>0$), eigenvalues and eigenvectors of the Hamiltonian $H$ are
$$
E_{+}=E+\sqrt{J^2+\Delta^2},\quad |\psi_{+}\rangle=\frac{1}{\sqrt{2\left(J^2+\Delta^2+\Delta\sqrt{J^2+\Delta^2}\right)}}\pmatrix{ \Delta+\sqrt{J^2+\Delta^2} \cr J },
$$
$$
E_{-}=E-\sqrt{J^2+\Delta^2},\quad |\psi_{-}\rangle=\frac{1}{\sqrt{2\left(J^2+\Delta^2-\Delta\sqrt{J^2+\Delta^2}\right)}}\pmatrix{ \Delta-\sqrt{J^2+\Delta^2} \cr J }.
$$

In the limit of large $\Delta$ (compared with $J$) we get
\begin{equation}\label{Delta+}
|\psi_{+}\rangle=\pmatrix{ 1 \cr 0 }=|1\rangle,\quad |\psi_{-}\rangle=\pmatrix{ 0 \cr 1 }=|2\rangle,\quad \Delta\to +\infty,
\end{equation}
\begin{equation}\label{Delta-}
|\psi_{+}\rangle=\pmatrix{ 0 \cr 1 }=|2\rangle,\quad |\psi_{-}\rangle=\pmatrix{ 1 \cr 0 }=|1\rangle,\quad \Delta\to -\infty,
\end{equation}
$$
|1\rangle=\frac{\Delta+\sqrt{J^2+\Delta^2}}{\sqrt{2\left(J^2+\Delta^2+\Delta\sqrt{J^2+\Delta^2}\right)}}|\psi_{+}\rangle + \frac{\Delta-\sqrt{J^2+\Delta^2}}{\sqrt{2\left(J^2+\Delta^2-\Delta\sqrt{J^2+\Delta^2}\right)}} |\psi_{-}\rangle,
$$
$$
|2\rangle=\frac{J}{\sqrt{2\left(J^2+\Delta^2+\Delta\sqrt{J^2+\Delta^2}\right)}}|\psi_{+}\rangle + \frac{J}{\sqrt{2\left(J^2+\Delta^2-\Delta\sqrt{J^2+\Delta^2}\right)}} |\psi_{-}\rangle.
$$

For $E_1\ne E_2$ evolution will not lead to a complete transition of the wave function --- probability if the state $|2\rangle$ (if we start from $|1\rangle$) will oscillate but will not be equal to one
$$
\langle 2|e^{-itH}|1\rangle=\frac{J}{2\sqrt{J^2+\Delta^2}}e^{-itE_{+}} \left(1-e^{2it\sqrt{J^2+\Delta^2}}\right).
$$

For low compared to the half-difference of energies $\Delta$ off-diagonal matrix elements $J$ the amplitude will be small. As a result the transition probability will be low --- already for $\Delta=J$  the transition probability is twice lower. The splitting of energy levels will grow to $2\sqrt{J^2+\Delta^2}$ and the beat period will decrease.

\medskip

\noindent{\bf Landau-Zener formula --- simple discussion}. Let us discuss the transitions for time-dependent Hamiltonian describing, for example, transitions between the electronic states of a molecule with nuclear vibrations (which provide the time dependence of the Hamiltonian).  Let us rewrite the formula (\ref{Amplitude0}) for the transition probability (the amplitude modulus squared) for transition $|1\rangle\to|1\rangle$  in the form
\begin{equation}\label{Probability}
|\langle 1|Te^{-i\int_0^t H(\tau) d\tau}|1\rangle|^2={\rm Tr}\,\left[ |1\rangle\langle 1| Te^{-i\int_0^t H(\tau) d\tau}|1\rangle\langle 1|Te^{i\int_0^t H(\tau) d\tau} \right],
\end{equation}
for a single transition one should consider $t=T$ (where $T$ is the transition period). To obtain the Landau--Zener formula, we consider the linear time-dependent Hamiltonian
\begin{equation}\label{HamiltonianLZ}
H(t)=\pmatrix{vt & J \cr J & ut \cr},
\end{equation}
Here the evolution operator is written in terms of the $T$-exponent (which for a time-independent Hamiltonian coincides with the usual exponent).

Diabatic energy levels (energy levels in absence of transitions $J=0$) have the form $E_1(t)=vt$, $E_2(t)=ut$, and adiabatic energy levels (energy levels of the total time-dependent Hamiltonian) are equal to
\begin{equation}\label{Epm}
E_{\pm}(t)=\frac{(v+u)t}{2}\pm\sqrt{\left(\frac{(v-u)t}{2}\right)^2+J^2}.
\end{equation}

Intersection point of diabatic energy levels $E_1(t)=E_2(t)$ is given by $t=0$ (the avoided crossing point for adiabatic energy levels). Intersection point of adiabatic energy levels $E_+(t)=E_-(t)$ is imaginary (modulus of this value can be considered as the half-width of the transition)
\begin{equation}\label{branching}
t=\pm \frac{2iJ}{v-u}.
\end{equation}

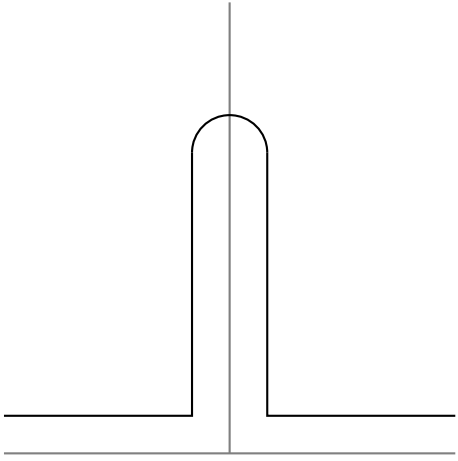
\begin{pspicture}(6,6)
\psline[linecolor=gray](0,0)(6,0)
\psline[linecolor=gray](3,0)(3,6)
\psline(0,0.5)(2.5,0.5)(2.5,4)
\psarc(3,4){0.5}{0}{180}
\psline(3.5,4)(3.5,0.5)(6,0.5)
\end{pspicture}

Let us consider the transition amplitude (taking in account that the adiabatic level for $t\to\pm\infty$ tends to different diabatic levels)
\begin{equation}\label{Amplitude}
\langle 1|Te^{-i\int_{-\infty}^{+\infty} H(\tau) d\tau}|1\rangle=\langle \psi_{+}(+\infty)|Te^{-i\int_{-\infty}^{+\infty} H(\tau) d\tau}|\psi_{-}(-\infty)\rangle.
\end{equation}
Let us approximate this amplitude by the expression
\begin{equation}\label{Amplitude1}
A=\langle \psi_{+}(+\infty)|e^{-i\int_{C}E_{\pm}(\tau) d\tau}|\psi_{-}(-\infty)\rangle,
\end{equation}
containing (instead of the integral of the Hamiltonian) the integral of the eigenvalue of the Hamiltonian $\int_{C}E_{\pm}(\tau) d\tau$ over contour $C$ in the complex plane of the form: it runs along the real axis from $-\infty$ to zero, then it runs along the imaginary axis to the point $\frac{2iJ}{v-u}$ and back to zero, then it runs along the real axis to $+\infty$. At the left half of the contour (up to point (\ref{branching})) we integrate the eigenvalue $E_{-}(\tau)$ and at the right half of the contour the eigenvalue $E_{+}(\tau)$ (in the branching point $\frac{2iJ}{v-u}$ these two eigenvalues coincide). It is easy to see that outside the neighborhood of zero and the imaginary axis contributions to integrals (\ref{Amplitude}) and (\ref{Amplitude1}) are almost equal (i.e. expression (\ref{Amplitude1}) approximates (\ref{Amplitude}) through the transition of adiabatic eigenvalues $E_{\pm}(\tau)$ one to the other through the complex plane). The integral along the real axis contributes to the amplitude (\ref{Amplitude1}) only oscillations. Nontrivial contribution to (\ref{Amplitude1}) gives the integral along the imaginary axis from zero to $\frac{2iJ}{v-u}$ and back, moreover, contributions from different sides of the loop have opposite signs (the direction of integration along the contour at different sides of the loop is opposite). We obtain for the integrand at the loop
$$
E_+(\tau)-E_-(\tau)=2\sqrt{\left(\frac{(v-u)\tau}{2}\right)^2+J^2}=2J\sqrt{1-z^2},\quad \tau=\frac{2iJ z}{v-u},
$$
and for the integral
\begin{equation}\label{integral}
\int_{0}^{t}(E_+(\tau)-E_-(\tau))d\tau=2J\frac{2iJ }{v-u}\int_0^1 \sqrt{1-z^2}dz=\frac{4iJ^2 }{v-u}\frac{\pi}{4}=\frac{i\pi J^2}{v-u}.
\end{equation}
The imaginary unit appeared due to the complex change of the variable of integration and it is responsible for the contribution with decreasing (not oscillations) in the integral, because in the formula for the transition amplitude (\ref{Amplitude1}) the integral is multiplied by one more imaginary unit, hence the decrease in the amplitude comes from the imaginary part of the integral. We obtain for the probability of transition between adiabatic levels, i.e. the square of the modulus of the amplitude (\ref{Amplitude}) approximation in the form of Landau--Zener formula
$$
P=|A|^2=\exp\left(-\frac{2\pi J^2}{\hbar|v-u|}\right).
$$

\section{Landau--Zener formula and master equations}

\noindent{\bf Non-secular dynamics}.
Let us consider the dynamics of density matrix with the time-dependent Hamiltonian
\begin{equation}\label{vonNeumann}
\frac{\partial}{\partial t}\rho(t)=-\frac{i}{\hbar}[H(t),\rho(t)],\quad H(t)=\pmatrix{{\bf v}\cdot{\bf q}(t) & J \cr J & {\bf u}\cdot {\bf q}(t) \cr}.
\end{equation}

The Hamiltonian (\ref{HamiltonianLZ}) can be considered as a linearization of above Hamiltonian in vicinity of the point of avoided crossing of adiabatic levels, i.e. the intersection point of diabatic levels ${\bf v}\cdot {\bf q}(t)={\bf u}\cdot {\bf q}(t)$ (here bold notation is used for vectors). According to Landau--Zener formula transitions between levels occur in vicinity of the avoided crossing point. If we choose ${\bf v}\cdot {\bf q}(t)=v\sin\omega t$, ${\bf u}\cdot {\bf q}(t)=u\sin\omega t$, avoided crossing points will satisfy $\sin\omega t=0$, $t=\pi k/\omega$. The physical meaning of this Hamiltonian is as follows: ${\bf q}(t)$ describes a vibron (some mode of vibrations of nuclei in molecules), transitions occur at the moments of passing the avoided crossing of energy levels, for the considered model it is $t=\pi k/\omega$, $k\in\mathbb{Z}$, the probability of transition between diabatic levels in one pass of the avoided crossing is
\begin{equation}\label{1-P}
1-P=1-\exp\left(-\frac{2\pi J^2}{\hbar\omega|v-u|}\right).
\end{equation}

Therefore the operation of the dynamics (\ref{vonNeumann}) looks as follows --- the vibron ${\bf q}(t)$ modulates transitions between two energy levels, the transitions are fast and outside the vicinity of the avoided crossing there is no nontrivial dynamics.

In the slow time scale $\tau=t/\omega$ (where the vibration period $T=2\pi/\omega<<1$) we will not observe the discussed above discrete dynamics but will observe some transition rate. In this time scale the dynamics of the system will be described by the von Neumann equation (where we put $\hbar=1$)
\begin{equation}\label{vonNeumann_slow}
\frac{\partial}{\partial \tau}\rho(\tau)=-i[H,\rho(\tau)],\quad H=s\pmatrix{0 & 1 \cr 1 & 0 \cr},
\end{equation}
with non-secular Hamiltonian, i.e. in the slow time scale Landau--Zener transitions became non-secular transitions. The amplitude $s$ of non-secular mode depends on the frequency of the vibron and the probability (\ref{1-P}) of Landau--Zener transition and increases with increasing of (\ref{1-P}).

In \cite{Chin2013} ''laser-like pumping'' of coherences in quantum photosynthesis is discussed, these vibronic coherences interact with excitons, causing Rabi oscillations for exciton transitions. Slowly decaying vibrational coherences maintain the coherence of excitons. This mechanism will be effective if the vibron energy coincides with the energy difference of the exciton levels, i.e. there is a resonance, moreover, this resonance is observed experimentally, which is mentioned in many works on quantum photosynthesis \cite{Kolli2012}, \cite{Novoderezhkin2015}, \cite{Novoderezhkin2016}, \cite{Novoderezhkin2019}, \cite{Detrapping}.

In the approach of this paper, this resonance can be interpreted as follows. Probability (\ref{1-P}) of transition between diabatic energy levels is large if velocity of the vibron at the moment of passing of the avoided crossing point is small. Thus the transfer efficiency can be increased by slowing down the vibron at the moment of passing through the avoided crossing. Let us consider for (\ref{vonNeumann}) the following ansatz for diabatic energy levels corresponding to vibron ${\bf q}(t)$:
\begin{equation}\label{1-cos}
{\bf v}\cdot {\bf q}(t)=v(1-\alpha\cos\omega t),\quad {\bf u}\cdot {\bf q}(t)=u(1-\alpha\cos\omega t),\quad \alpha>0,\quad u\ne v .
\end{equation}
Here $u$, $v$ are called sweep velocities.

The avoided crossing is given by the equation
$$
\alpha\cos\omega t=1.
$$

For this ansatz for $\alpha<1$ there are no avoiding crossing points and transitions are effectively suppressed; for $\alpha=1$ the vibron passes the avoided crossing at zero speed, which gives the maximum efficiency for the transfer of excitons between diabatic energy levels; for $\alpha>1$ velocity of the vibron at avoided crossing points increases and the transition probability (\ref{1-P}) decreases. With this choice of vibron for $\alpha=1$ the transition is adiabatic, its energy coincides with the difference in energies of diabatic levels, i.e. a resonance is observed. Thus, the proposed model of the vibronic mechanism of exciton transfer (\ref{vonNeumann}), (\ref{vonNeumann_slow}), (\ref{1-cos}) explains the experimentally observed resonance between vibrons and excitons.

In more details: energy difference for adiabatic levels for $0\le\alpha\le 1$ oscillates between
$$
\sqrt{4J^2+[(v-u)(1-\alpha)]^2} \le E_{+}(t)-E_{-}(t) \le  \sqrt{4J^2+[(v-u)(1+\alpha)]^2}.
$$
For the resonant case $\alpha=1$ this reduces to
$$
2J \le E_{+}(t)-E_{-}(t) \le  2\sqrt{J^2+[v-u]^2}.
$$

Here $\cos\omega t=1$, $t=2\pi n/\omega$, $n\in \mathbb{Z}$ is the avoided crossing for the resonant case. In this case the difference of diabatic energy levels oscillates between 0 and $2|v-u|$  and for adiabatic levels the energy difference oscillates between $2J$ and $2\sqrt{J^2+[v-u]^2}$. In average the (diabatic) energy difference is $|v-u|$ which can be considered as the vibron amplitude (it is exactly the amplitude if one of sweep velocities equals zero, say $u=0$). Let us stress that here we discuss the resonance in energies of the vibron (which depends on the amplitude $\alpha$ of the vibrations in (\ref{1-cos})) and the energy difference between exciton states (i.e. for this resonance not the frequency but the amplitude of the vibron is important).

Modification of the parameters of the vibron (in particular the amplitude) can be achieved by modifying the proteins in vicinity of chromophores.
It was mentioned \cite{Novoderezhkin2019} that the mechanism of charge separation and transfer in photosynthetic reaction centers is ''fragile'' --- mutations leading to a slight change in the spatial conformation of proteins in the vicinity of the reaction center can reduce the rate of exciton current by two orders of magnitude. From the point of view of the model (\ref{1-cos}) this behavior can be easily explained --- in the resonant case (when the vibron passes the avoided crossing at a low speed), the variations in the vibron parameters ${\bf q}(t)$ can result in the fact that the vibron will ''miss'' the avoided crossing and the Landau--Zener transitions will be ineffective (the system will remain at the same diabatic level). For example, one can choose $|\alpha|<1$ in (\ref{1-cos}) to make the avoided crossing point non-existent.

To describe decoherence and dissipation during the transition between electronic states one has to add to the right-hand side of the equation (\ref{vonNeumann_slow}) a dissipative term
\begin{equation}\label{theta}
\theta(\rho)=
\gamma^{-}
\left(
\langle 2|\rho|2\rangle |1\rangle\langle 1|
-{1\over 2}
\{\rho,|2\rangle\langle 2|\}\right)+
\gamma^{+}
\left(\langle 1| \rho |1\rangle |2\rangle\langle 2|
-{1\over 2}
\{\rho,|1\rangle\langle 1| \}\right),
\end{equation}
where $\gamma^{+}/\gamma^{-}=e^{-\beta(\varepsilon_2-\varepsilon_1)}$ and $\varepsilon_2$, $\varepsilon_1$ are energies of levels $|2\rangle$, $|1\rangle$ correspondingly, $\varepsilon_2>\varepsilon_1$ and $\beta$ is the inverse temperature for this transition (for example $\beta^{-1}=300$K for room temperature).

Thus, we obtain a dynamics of the form
\begin{equation}\label{vonNeumann_slow1}
\frac{\partial}{\partial \tau}\rho(\tau)=-i[H,\rho(\tau)]+\theta(\rho),
\end{equation}
analogous to considered in \cite{Chin2013}. Dissipative generator $\theta$ in this equation generates quantum transfer, the non-secular term $[H,\cdot]$ operates as a quantum transfer amplifier. This model is analogous to the model of laser, the non-secular term is analogous to the laser mode (see \cite{Haken} for theory of lasers). Laser-like model of quantum amplification of transfer in quantum photosynthesis was discussed in \cite{laser}. Namely it was shown that the current of excitons described by equation of the kind (\ref{vonNeumann_slow1}) in the case of presence of vibrons (with a non-secular term in the master equation) increases as
$$
F\mapsto F+ s^2 F_1,
$$
where $s$ is the parameter in (\ref{vonNeumann_slow}) (non-secular ''laser'' mode amplitude).

In contrast to the laser model, the non-secular term here arises not due to pumping, but according to the semiclassical Franck--Condon principle. By this principle, upon excitation of electronic degrees of freedom, nuclear oscillations (vibrons) are simultaneously excited, because if electronic states are excited, the states of nuclei become nonequilibrium.

In the literature on quantum photosynthesis, see for example  \cite{Novoderezhkin2016}, \cite{Kolli2012}, \cite{Novoderezhkin2015}, \cite{Novoderezhkin2019}, \cite{Detrapping}, \cite{Stark} the presence of resonances  between energies of vibrons and energy differences of exciton states coupled with these vibrons is widely discussed. In accordance with the discussion of the present paper, this resonance is important for efficient operation of transitions between electron energy levels  by the Landau--Zener mechanism (i.e. to increase the parameter $s$ in (\ref{vonNeumann_slow})). Application of the Franck--Condon principle of simultaneous generation of excitons and vibrons within the framework of uniform eigenvectors of the Hamiltonian was discussed in \cite{Novoderezhkin2015}, also the polaron master equations \cite{Kolli2011}, \cite{Kolli2012} and equations with non-secular terms \cite{Chin2013} were discussed. The approach of the present paper allows to give a simple description of non-secular terms in master equations which amplify quantum transfer. Also the resonance condition for the vibron energy and the energy difference of exciton states paired to the vibron is explained.

\end{document}